# Critical behavior at superconductor-insulator phase transitions near one dimension

Igor F. Herbut

Department of Physics and Astronomy, University of British Columbia, 6224 Agricultural Road, Vancouver B. C., Canada V6T 1Z1

**Abstract:** I argue that the system of interacting bosons at zero temperature and in random external potential possesses a simple critical point which describes the proliferation of disorder-induced defects in the superfluid ground state, and which is located at weak disorder close to and above one dimension. This makes it possible to address the critical behavior at the superfluid-Bose glass transition in dirty boson systems by expanding around the lower critical dimension of $d = 1$. Within the formulated renormalization procedure near $d = 1$ the dynamical critical exponent is obtained exactly and the correlation length critical exponent is calculated as a Laurent series in the parameter $\sqrt{\epsilon}$, with $\epsilon = d - 1$: $z = d$, $\nu = 1/\sqrt{3\epsilon} + O(\sqrt{\epsilon})$ for the short-range, and $z = 1$, $\nu = \sqrt{2/3\epsilon} + O(\sqrt{\epsilon})$, for the long-range Coulomb interaction between bosons. The identified critical point should be stable against the residual perturbations in the effective action for the superfluid, at least in dimensions $1 \leq d \leq 2$, for both short-range and Coulomb interactions. For the superfluid-Mott insulator transition in the system in a periodic potential and at a commensurate density of bosons I find $\nu = (1/2\sqrt{\epsilon}) + 1/4 + O(\sqrt{\epsilon})$, which yields a result reasonable close to the known XY critical exponent in $d = 2$. The critical behavior of the superfluid density, phonon velocity and the compressibility in the system with short-range interactions is discussed.

# 1   Introduction

The transition between the superfluid (SF) and the localized insulating (Bose-glass, BG) ground states in the disorder bosonic systems at zero temperature is a paradigmatic example of a quantum phase transition [1], [2]. It is believed to be relevant for a number of experimentally observed superconductor-insulator transitions at low temperatures, including those in Josephson junction arrays [3], amorphous superconducting films [4], underdoped high-$T_c$ cuprates [5], $^4He$ and $^3He$ in disordered media [6], [7], and possibly in two-dimensional electron gas in Si-MOSFETs [8]. The vortex transitions in type-II superconductors in magnetic field in presence of correlated disorder have also been related to this problem [9], [10]. Despite of its ubiquity in nature and good qualitative understanding of the insulating phase [2], the critical behavior at the SF-BG transition is only poorly known. This may appear surprising, particularly when contrasted with the greatly successful theory of classical critical phenomena [11], even in the presence of quenched disorder [12]. The source of the difficulty can be traced to the difference between the classical and the quantum disordered systems that becomes most apparent in the path integral formulation of the quantum problem: $d$-dimensional quantum system at $T = 0$ maps onto a $d+1$-dimensional classical theory, but the disorder potential appears random only in $d$-dimensions, while it is completely correlated in the remaining dimension which represents the imaginary time. This is because static disorder is represented by a time-independent random potential. The unpleasant consequence of this fact is that there is no upper critical dimension for the SF-BG transition in the standard sense. One way to try to deal with this problem is to demote a number of imaginary-time dimensions from unity to a small number $\epsilon_\tau$, and then to perform a double expansion in both $\epsilon_\tau$ and $\epsilon = 4 - d - \epsilon_\tau$ [13]. Unfortunately, the perturbative results for the critical exponents obtained this way appear to have a very poor convergence in $\epsilon_\tau$, rendering this approach of limited use for calculations of the critical exponents. That this may not be a coincidence could be suspected on the basis of certain exact inequalities which should be satisfied by the exponents at the disordered critical points [14], and which dictate that their values should be very different than at the Gaussian fixed point. For example, in $d = 2$ and $\epsilon_\tau = 1$, at the SF-BG critical point one expects $\nu \geq 1$, and $z \approx 2$, in contrast to $\nu = 1/2$ and $z = 1$ at the Gaussian fixed point for the relativistic bosons. In fact, within the double $\epsilon$-expansion the disordered and the Gaussian fixed points generally do not coalesce at the upper critical dimension $d_{up} = 4$, but instead the disordered critical point stays always non-trivial (except when $\epsilon = \epsilon_\tau = 0$), and becomes unstable only in dimensions higher than some $d_c > d_{up}$ [15]. All this is of course completely different from the standard $4 - \epsilon$ expansion for the classical systems [11], where the Wilson-Fisher critical point coincides with the Gaussian fixed point at the upper critical dimension, becomes unstable right above it, and the critical exponents in the physical dimensions are numerically not too different from their Gaussian values. One might speculate that the failure of the double $\epsilon$-expansion may be due to non-analyticity of the critical exponents as functions of $\epsilon_\tau$ [16].

Recognizing the limitations of the double $\epsilon$-expansion many authors turned to different ways of addressing the SF-BG transition: large-N [17], real-space methods [18], strong-coupling expansion [19], numerical calculations [20], and the renormalization group in fixed dimension [21] have all been applied to this problem. In absence of a small parameter one is left feeling uneasy with the results obtained by these methods, and it would still be desirable to have a controlled analytical approach to the SF-BG critical point which would be free of the conceptual and the calculational problems of the double $\epsilon$-expansion. Besides providing a practical tool for calculation of the universal quantities at the SF-BG phase transition, such a scheme would be expected to be able to correctly discern different universality classes of superconductor-insulator transitions, similarly to the celebrated $\epsilon$-expansion in the field of classical critical phenomena.

In this paper I present an attempt to develop such a systematic theory of the SF-BG critical point as a perturbative expansion in deviation of spatial dimensionality $d$ of the quantum system from unity. That the parameter $\epsilon = d - 1$ effectively controls the value of disorder at the superconductor-insulator critical point has been suggested previously by the author in [22], where this was used to study the superconductor-Anderson insulator transition in the model of disordered attracting spinless fermions near one dimension. The basic idea is rather general and may be understood qualitatively as follows. Consider a quantum system of interacting bosons (which may be Cooper pair composites, for instance) in $d = 1$ and at $T = 0$: in absence of any external potential the Bose single-particle operator will exhibit a power-law quasi-long-range correlations, with a non-universal power $K$ depending on the microscopic interactions. $K$ may therefore be understood as an exactly marginal coupling constant. What will be the effect of a random external potential? It must suppress superfluidity by trying to localize the elementary excitations. In the language of renormalization group this means that it should produce a *positive* term in the recursion relation for $K$, in attempt to drive the Bose single-particle correlation function shorter ranged. The renormalization group flow of the disorder coupling constant $D$ will in turn be affected by $K$, so that if $K$ is larger than some $K^*$, small $D$ is relevant and vice versa [23], [24]. The crucial point is that the marginality of $K$ in $d = 1$ without disorder forces the SF-BG fixed point to lay at $D^* = 0$ (and some finite $K = K^*$, see Figure 1.b ) in $d = 1$. To use this fact to develop an expansion in small parameter $\epsilon = d - 1$ for the problem with disorder one must also observe the following: power-law superfluid correlations (without disorder) are characteristic of $d = 1$ [25]; in $d > 1$, true long range order will develop at $T = 0$. In $d = 1 + \epsilon$ one may still think of $K$, but no longer as a marginal coupling; it should slowly (controlled by $\epsilon$) scale to zero, so that the single-particle correlation function tends to a finite constant at large distances. Thus a small deviation of dimensionality of the system from unity should produce a small *negative* term in the recursion relation for $K$, which then may be balanced by the positive term induced by disorder. This opens a possibility of having a random critical point in the physical region of the coupling space which is infinitesimal in disorder slightly above one dimension (see Figure 1).

The central assumption made here is that the SF-BG critical point in $d = 2$ smoothly evolved from the trivial (in the sense that $D^* = 0$) SF-BG critical point in $d = 1$. The additional motivation for this idea comes from the study of the disordered Bose-Hubbard model [21]. Using the duality transformations to map the dirty-boson model on lattice onto a field theory for defect degrees of freedom, it was noticed that the SF-BG critical point in the dual formulation arises through similar renormalization scenarios in $d = 1$ and $d = 2$. The form of the dual theories in one and two dimensions is quite different [21], reflecting more complicated nature of defects in higher dimensions, nevertheless, in both cases it is the coupling of the external potential to the field that describes topological defects of the phase of the superfluid order parameter which ultimately determines the critical behavior. The physical mechanism for destruction of superfluidity at $T = 0$ is postulated to be the same in $d = 1$ and $d = 2$: the unbinding of phase slip-antislip (in d=1) or vortex-antivortex (in $d = 2$) pairs *induced* by disorder destroys the phase coherence. The critical point which describes this transition is strongly coupled in $d = 2$, but becomes weakly coupled in $d = 1$, which suggests an expansion around the latter solvable case.

In the rest of the paper this idea is used to calculate the critical exponents at the SF-BG transition in the systems of disordered bosons with short-range and long-range Coulomb repulsion *near* one dimension. My starting point is the effective low-energy action for the phonon excitations above the SF ground state [26] in presence of an external random potential. Near $d = 1$ the non-trivial part of the recursion relations is determined from the theory as precisely in $d = 1$, while the effect of dimensionality enters through the non-vanishing canonical dimensions of the coupling constants. To handle the disorder term I take advantage of the density representations of the theory [27] in $d = 1$, where it may be understood as dual to the phase representation [21], and in which the external potential couples to the operator that creates $2\pi$ phase slips in the superfluid ground state. I first study the case of the short-range interaction between bosons in detail, later to extend the calculation onto the Coulomb universality class. The recursion relations are derived to $O(\epsilon^2)$ and the correlation length critical exponents $\nu$ and the anomalous dimension $\eta$ are calculated as series in $\epsilon$ to two lowest orders, whereas the dynamical critical exponent is obtained exactly. In particular, within the presented formalism the conjecture of Fisher et al. [2], [28] that the dynamical exponent $z = d$ and $z = 1$ for the short-range and the Coulomb repulsion between bosons, respectively, holds exactly as long as the critical point is smoothly connected to the one in $d = 1$. For the short-range interactions, both the superfluid density and the phonon velocity renormalize to zero at the SF-BG transition, as power-laws with powers that vanish as $\sim \sqrt{\epsilon}$ close to $d = 1$. Compressibility on the other hand approaches a finite constant at the transition, in accordance with the expected gapless nature of the Bose glass phase.

The paper is organized as follows. In the next section I define the problem and determine the canonical dimensions of the coupling constants in the effective action in the case of short-range interactions. In section 3 the recursion relations for the coupling constants to two lowest orders in $\epsilon = d - 1$ are presented, and the critical exponents for the case of short-range repulsion are obtained. In section 4 I examine the relevance of the residual terms in the superfluid effective action at the SF-BG critical point. Section 5 presents the results for the critical exponents in the case of Coulomb repulsion between the bosons. I test the expansion method against the known results for the XY universality class relevant for the superfluid-Mott insulator transition in a periodic potential in section 6. Summary and the discussion of the main points is given in the last section, and the calculational details are relegated to the appendices.

## 2 Hamiltonian and the canonical dimensions

I begin by defining the system of interacting bosons with the standard second-quantized Hamiltonian:

$$H = \frac{\hbar^2}{2m} \int d^d\vec{x} |\nabla \Psi(\vec{x})|^2 + \frac{1}{2} \int d^d\vec{x} d^d\vec{y} v(\vec{x} - \vec{y}) |\Psi(\vec{x})|^2 |\Psi(\vec{y})|^2, \quad (1)$$

where $v(\vec{r})$ is a short-range repulsive interaction, with $\tilde{v}(\vec{q}=0)$ finite, $|\Psi|^2 = \Psi^\dagger \Psi$, and $\Psi$ satisfies standard bosonic commutation relations. The Hamiltonian (1) has a superfluid ground state in all $d \geq 1$. The form of the effective theory for the low-energy excitations may be derived by writing the Bose operator in the density-phase representation as $\Psi(\vec{x}) = \rho^{1/2}(\vec{x}) \exp i\phi(\vec{x})$ [26], where $\rho(\vec{x}) = \rho_0 + \Pi(\vec{x})$, with $\Pi(\vec{x})$ being a local deviation from the average density of bosons $\rho_0$. Expanding the Hamiltonian to the quadratic order in $\Pi$ and $\phi$, the imaginary time quantum mechanical action $S = \int d^d\vec{x} d\tau (\hbar \Psi^\dagger \partial_\tau \Psi + H[\Psi^\dagger, \Psi])/\hbar$ at finite temperature $T$ becomes:

$$S_{\phi,\Pi} = \int d^d\vec{x} \int_0^\beta d\tau (\frac{v_J}{2\pi} (\nabla \phi(\vec{x}, \tau))^2 + 2\pi v_N \Pi^2(\vec{x}, \tau) + i\Pi(\vec{x}, \tau) \partial_\tau \phi(\vec{x}, \tau) + ...), \quad (2)$$

where $\beta = \hbar/k_B T$, and the terms of higher order in $\Pi(\vec{x})$ and $\phi(\vec{x})$ and their derivatives have been omitted for the moment. The boundary conditions satisfied by the fluctuating fields are $\phi(\vec{x}, \beta) = \phi(\vec{x}, 0) + 2\pi n(\vec{x})$, with $n(\vec{x})$ integer, and $\Pi(\vec{x}, \beta) = \Pi(\vec{x}, 0)$. At $T = 0$, the action in Eq. (2) may be understood as an effective low-energy theory for the superfluid by defining the coupling constants as $v_J = \hbar \pi \rho_s / m$ and $v_N = 1/(4\pi\hbar\kappa)$, where $\rho_s$ is the superfluid density and $\kappa$ is the compressibility of the system. The effective theory $S_{\phi,\Pi}$ then represents an exact description of the low-momentum ($|\vec{k}| < \Lambda$, where $\Lambda << 2\pi\rho_0$ is an ultraviolet cutoff) phonons, which are the Goldstone modes of the broken $U(1)$ symmetry in the superfluid ground state. It is just the Landau quantum hydrodynamics for the superfluid $^4He$ [29]. The presence of the last imaginary term in Eq. (2) implies that in the subspace of low-energy states the density and the phase should be considered as canonically conjugate variables:

$$[\Pi(\vec{x}), \phi(\vec{y})] = -i\delta^d(\vec{x} - \vec{y}). \quad (3)$$

The terms of higher order in $\Pi$ and $\phi$ omitted in Eq. (2) describe the residual interactions between the phonons and the deviation of phonon dispersion from linearity. They will be considered in Section 4, where I will argue that they are irrelevant at the SF-BG critical point, at least for $1 \leq d \leq 2$. Note that these perturbations lead to infrared-finite corrections to the low-energy theory for the superfluid, which I will assume are already included in definitions of the coupling constants $v_J$ and $v_N$.

Let me determine the canonical dimensions of the coupling constants $v_J$ and $v_N$ from the effective theory (2) for the non-interacting excitations at $T = 0$. Anticipating the effects of disorder, consider the change of these couplings under rescaling of lengths, but allowing for the dimension of the imaginary time to be $[\tau] = -z$ (that is $\tau \sim L^z$, where $L$ is a length), with $z$ as an undetermined dynamical exponent. The canonical dimensions are

$$[v_J] = z + d - 2, \quad [v_N] = z - d, \tag{4}$$

as follows from recalling that $\phi(\vec{x})$ is a phase, so $[\phi] = 0$, and $[\Pi] = d$. The reader recognizes the first equation as essentially the Josephson scaling relation for the superfluid density, $\rho_s \sim \xi^{2-d-z}$ [1], while the second one is the analogous relation for the compressibility, $\kappa \sim \xi^{z-d}$ [2]. $\xi$ is the diverging correlation length at the transition. Note that the canonical dimensions of both couplings vanish in d=1, since for the non-interacting excitations $z = 1$. The last statement is equivalent to the algebraic decay of the single-particle correlation function in $d = 1$ and at $T = 0$ at large distances, or, more precisely

$$\langle \Psi^+(\vec{x})\Psi(0) \rangle \sim \rho_0 |\vec{x}|^{-(v_N/v_J)^{1/2}} \tag{5}$$

when $|\vec{x}| \to \infty$. In $d > 1$ the single-particle correlation function in Eq. (5) tends to a finite constant at large separations.

Consider adding a term with an external potential to the Hamiltonian (1):

$$H_r = \int d^d \vec{x} V(\vec{x}) |\Psi(\vec{x})|^2, \tag{6}$$

where $V(\vec{x})$ is a random function of the coordinate. Evidently, the external potential couples directly only to the particle density and not to the phase. One may still perform the Gaussian integration over $\Pi$ in the action (2), to obtain a complex effective action for the phase $\phi$, with disorder entering via an awkward imaginary term $i \int V(\vec{x}) \partial_\tau \phi(\vec{x}, \tau)$. Since disorder is static, after integration over imaginary time one finds that this gives a purely boundary contribution, in which the external potential couples to the windings ($n(\vec{x})$) of the phase in the imaginary time. Fortunately, in dimensions $d = 1$ and $d = 2$ there exists a real (dual) representation of the effective action in terms of the particle density [21], [30], [31], in which disorder appears in a less forbidding form. Here I follow Haldane [27], and in $d = 1$ introduce a new field $\theta'(x, \tau)$ in the action (2) as $\Pi(x, \tau) = \pi^{-1} \partial_x \theta'(x, \tau)$. Performing the average over disorder using replicas [32], after integration over the superfluid phase the effective replicated low-energy theory in $d = 1$ becomes

$$S_\theta = K \sum_{i=1}^N \int dx \int_0^\beta d\tau (c^2(\partial_x \theta_i)^2 + (\partial_\tau \theta_i)^2) - D\rho_0^2 \sum_{i,j=1}^N \int dx d\tau d\tau' \cos 2(\theta_i(x,\tau) - \theta_j(x,\tau')), \tag{7}$$

where

$$\theta_i(x, \tau) = \theta'_i(x, \tau) + \frac{1}{4v_N} \int_{-\infty}^x V(z) dz, \tag{8}$$

and I introduced the standard combinations of the coupling constants $K = 1/(2\pi v_J)$, $c^2 = 4v_N v_J$ for later convenience. $c$ represents the velocity of the phonon excitations, and $K$ is inversely proportional to the superfluid density. The limit $N \to 0$ at the end of calculations is assumed. Haldane's heuristic derivation of the effective action $S_\theta$ for bosons in one dimension [24], [27] is presented in Appendix A. It is worth mentioning that the action in Eq. (7) also arises as the long-distance theory of the disordered Bose-Hubbard model in one dimension [21], [30]. Disorder produces an effective interaction between replicas, which in $d = 1$ is reminiscent of the interaction term in the sine-Gordon theory. There are two important differences however. First, the interaction is non-local in imaginary time, which is a consequence of the quantum nature of the problem, and breaks the relativistic invariance of the pure system. Second, the interaction term is invariant under a shift of all fields by an arbitrary function of the coordinate $x$. Both features will have important consequences for the critical behavior at the SF-BG phase transition.

## 3   Recursion relations and the critical exponents

In the rest of the paper I will be interested in the Bose system at zero temperature, and therefore set $\beta = \infty$ in Eq. (7). Without disorder, the effective action in Eq. (2) describes the non-interacting excitations. Under the change of the cutoff $\Lambda \to \Lambda/s$, or equivalently under the rescaling of coordinates as $\vec{x} \to s\vec{x}$ and $\tau \to s^z \tau$, coupling constants $K$ and $c$ in general scale as determined by their relations to $v_J$ and $v_N$ and their canonical dimensions in Eqs. (4), with $z = 1$. Disorder introduces effective interactions between the excitations in any dimension, and in $d = 1$ the most relevant one has a simple form given by Eq. (7). Also, precisely in $d = 1$, both $K$ and $c$ have their canonical dimensions vanishing. Thus the strategy will be to calculate the recursion relations for $K$ and $c$ in $d = 1$ perturbatively in disorder from the action (7), and then to account for the effects of dimensionality when $d > 1$ by adding terms with the canonical dimensions and with a general $z$. The reader will note that this is precisely the logic of the minimal substraction scheme when used together with dimensional regularization in the theory of classical critical phenomena. Finally, I will determine the value of the dynamical exponent by requiring the existence of the fixed point of the recursion relations in $d > 1$.

I now proceed with the implementation of this program, and in $d = 1$ integrate out the modes of the field $\theta(x, \tau)$ which have the momentum $\Lambda > |k| > \Lambda/s$, for $\ln(s)$ infinitesimal, and with any imaginary frequency $-\infty < \omega < \infty$. The details of the perturbative calculation to two lowest orders in disorder are presented in Appendix B. Including the $d$-dependent canonical dimensions from the previous section and redefining the couplings as $\pi K \to K$, $\pi D \rho_0^2 / \Lambda^3 \to D$, I find the recursion relations:

$$\frac{dK}{d\ln(s)} = (2 - z - d)K + \frac{8D}{Kc^4} + O(D^2), \qquad (9)$$

$$\frac{dc}{d\ln(s)} = (z - 1 - \frac{4D}{K^2 c^4})c + O(D^2), \qquad (10)$$

$$\frac{dD}{d\ln(s)} = (d + 2z - \frac{1}{Kc})D + O(D^3). \qquad (11)$$

The last three equations represent the central result of this paper, and some remarks may be in order. To the second order in disorder $D$ the coefficient in front of $(\partial_\tau \theta)^2$ in Eq. (7) becomes renormalized by the integration over the fast modes, whereas the one in front of $(\partial_x \theta)^2$, which is the inverse of compressibility, does not. The recursion relation for the velocity $c$ is thus completely determined by the one for $K$, to two leading orders. To complete the scaling transformation I need to specify the dynamical exponent: I choose $z$ to keep the velocity $c$ constant under renormalization. This is essential to find a fixed point in the theory and also conveniently decouples the recursion relation for $c$ from the other two. Numerical value of $c$ is then completely arbitrary. Setting $c = 1$ and inserting $z = 1 + 4D/K^2 + O(D^2)$ into Eqs. (9) and (11), I find the fixed point of the above renormalization group equations at

$$K^* = \frac{1}{3} - \frac{\epsilon}{3} + O(\epsilon^2), \tag{12}$$

$$D^* = \frac{\epsilon}{36} + O(\epsilon^2). \tag{13}$$

At the fixed point the dynamical exponent thus equals

$$z = 1 + \epsilon + O(\epsilon^3). \tag{14}$$

Note that there is no $O(\epsilon^2)$ correction in the last result. The reason is the absence of renormalization of the coefficient in front of $(\partial_x \theta)^2$-term in the action (7) to the second order in disorder. In Appendix C I show that the exact symmetry of the disorder term in the one-dimensional theory in Eq. (7) under $\theta_i(x, \tau) \to \theta_i(x, \tau) + f(x)$ for arbitrary $f(x)$, guarantees that this coefficient remains unrenormalized to *all* orders in $D$ in $d = 1$. Within the proposed calculational scheme where the non-trivial part of the recursion relations follows from the theory as in $d = 1$, this represents the sufficient condition for that all higher corrections to $z$ in Eq. (14) vanish, i. e.

$$z = d \tag{15}$$

exactly. The conjecture of Fisher et al. [2] thus follows as an exact consequence of the formulated renormalization procedure near one dimension.

Linearizing the renormalization group flow around the critical point in Eqs. (12)-(13) (Figure 1) one finds the correlation length exponent which determines the flow along the relevant direction to be

$$\nu = \frac{1}{\sqrt{3\epsilon}} + O(\sqrt{\epsilon}). \tag{16}$$

The other eigenvalue which determines the flow along the SF-BG separatrix gives the leading correction to scaling close to $d = 1$. $\nu$ turns out to be a Laurent series in $\sqrt{\epsilon}$ and therefore of weaker dependence on dimensionality than one would expect. For $\epsilon = 1$ the result (16) yields $\nu \approx 0.58$, smaller than in the Monte Carlo calculations of Wallin et al. [20], where $\nu = 0.9 \pm 0.1$. The generalized Harris criterion [14] which requires that $\nu > 2/d$ at a disordered fixed point is also not satisfied at this order. It is possible that the higher order terms in the expansion will increase the value of $\nu$, and remedy both discrepancies. The divergence of $\nu$ as $\epsilon \to 0$ is a sign of vicinity of the lower critical dimension for the SF-BG transition.

Right at the critical point in $d > 1$ the average single-particle correlation function shows an algebraic decay at large distances (see Appendix D):

$$\overline{G(\vec{x})} \sim \rho_0 |\vec{x}|^{-K^* c^*}. \tag{17}$$

The anomalous dimension $\eta$ is defined by $K^* c^* = d + z - 2 + \eta$ [2], so that:

$$\eta = \frac{1}{3} - \frac{7\epsilon}{3} + O(\epsilon^2). \tag{18}$$

For $\epsilon = 1$ the above result yields a negative $\eta$, as one would expect based on the assumption that the single-particle density of states $N(e) \propto e^{\frac{d-2+\eta}{z}}$ at the critical point diverges at zero energy [2]. The result however is much more negative than the one obtained by Wallin et al in their Monte Carlo calculation [20]: $\eta = -0.1 \pm 0.15$. Higher order terms may bring these two estimates closer, although expansions of the anomalous dimension around the lower critical dimensions are usually notoriously badly convergent.

The reader may have noticed that the coefficient in front of the second term in the scaling equation for $K$ depends on the precise definition of the coupling $D$ and is therefore arbitrary. It is reassuring to realize that none of the exponents in fact depends on the value of this coefficient.

Upon approaching the SF-BG transition from the superfluid side, the velocity of phonons and the superfluid density in $d = 1 + \epsilon$ for $\epsilon$ small both vanish as

$$c \sim \delta^{\sqrt{\epsilon/3}}, \tag{19}$$

and

$$\rho_s \sim \delta^{2\sqrt{\epsilon/3}}, \tag{20}$$

where $\delta$ is a parameter that tunes through the transition. In contrast, in $d = 1$ both quantities approach finite values at the superfluid side of the transition, and then vanish discontinuously in the insulating phase. Compressibility $\kappa \sim \rho_s / c^2$, on the other hand, remains finite at the transition in all dimensions, as implied by the found exact relation $z = d$. Although finite, at the transition its value remains non-universal and dependent on the microscopic couplings in the system.

## 4 Residual terms in the effective action

For the fixed point in Eqs. (12)-(13) to govern the critical behavior at the SF-BG transition in $d > 1$ one must assure that the terms of higher order in $\Pi(\vec{x})$ and $\phi(\vec{x})$ and their derivatives which were dropped in the effective action $S_{\phi,\Pi}$ are irrelevant perturbations. These terms originate from the kinetic energy part of the Hamiltonian (1), which in the density-phase representation takes the form:

$$H_k = \frac{\hbar^2}{2m} \int d^d\vec{x}[\rho_o(\nabla\phi(\vec{x}))^2 + \Pi(\vec{x})(\nabla\phi(\vec{x}))^2 + \frac{(\nabla\Pi(\vec{x}))^2}{4(\rho_o + \Pi(\vec{x}))}]. \tag{21}$$

The effective action for the superfluid will thus contain additional parts coming from the second and the third term in the above expression:

$$S'_{\phi,\Pi} = \int d^d\vec{x}d\tau[v_x\Pi(\vec{x})(\nabla\phi(\vec{x}))^2 + v_y(\nabla\Pi(\vec{x}))^2 + O(\Pi(\nabla\Pi)^2)], \tag{22}$$

with $v_x$ and $v_y$ as two new coupling constants, which together with $S_{\phi,\Pi}$ in Eq. (2) determine the full effective low-energy action for the elementary excitations, without disorder.

To examine the relevance of the two new terms in $S'_{\phi,\Pi}$ at the SF-BG critical point in $d > 1$ we first need the canonical dimensions of $v_x$ and $v_y$ from Eq. (22): $[v_x] = z - 2$ and $[v_y] = z - 2 - d$. As long as $z = d$ at the SF-BG critical point, a small $v_y$ which describes the deviation of the spectrum from linearity will stay irrelevant close to $d = 1$, where its dimension may change form the canonical one only by $\epsilon$. It thus seems reasonable that this will remain so in physical dimensions $d = 2$ and $d = 3$. The coupling constant $v_x$ on the other hand, under the assumption that $z = d$ at the SF-BG fixed point is irrelevant by power-counting close to $d = 1$, but becomes marginal in $d = 2$ and relevant above $d = 2$. The situation is analogous to the effect of the $\psi^6$ term in the Ginzburg-Landau-Wilson theory at the Gaussian fixed point, and one must examine the corrections to the dimension of $v_x$ introduced by disorder. Dropping the $v_y$-term in Eq. (22), after integration over the phase in $d = 1$ one finds a term

$$S'_\theta = 4\pi v_x K^2 \int dx d\tau (\partial_\tau\theta)^2\{(\partial_x\theta) + O((\partial_x\theta)^2)\}, \tag{23}$$

in addition to $S_\theta$ in Eq. (7). To the two lowest orders in disorder I found previously that only the coupling $K$ in front of $(\partial_\tau\theta)^2$-term becomes renormalized. Under the change of cutoff $\Lambda \to \Lambda/s$ then

$$v_x K^2 \to v_x(s)K^2(s) = v_x K^2 + O(D^2, v_x^2). \tag{24}$$

Inserting $K(s)$ I then find the recursion relation for $v_x$:

$$\frac{dv_x}{d\ln(s)} = (z - 2 - \frac{16D}{K^2 c^4} + O(D^2))v_x + O(v_x^2). \tag{25}$$

When $d = 2$ and $z = 2$, at the fixed point $D^*/K^{*2} \sim 1$, so to truly answer the question of relevance of small $v_x$ at the SF-BG critical point we need all the higher order terms in $D$ in the bracket in the equation above. However, the first term in the expansion of the scaling dimension of $v_x$ in $D$ suggests that $v_x$ most likely stays irrelevant even for $d = 2$. The negative sign of this term comes from the effective increase of $K$ under the influence of disorder, which on physical grounds may be expected to persist to higher orders. It is plausible then that $v_x$ is irrelevant at the SF-BG critical point, at least for $1 \leq d \leq 2$.

Finally, consider the higher harmonics in the effective action in Eq. (7) that arise from the components of the random potential with the wave-vectors $k \sim 2m\pi\rho_0$ with $m \geq 2$ (see Appendix A):

$$S_{\theta,hh} = -\sum_{m=2}^{\infty} D_m \sum_{i,j=1}^{N} \int dx d\tau d\tau' \cos 2m(\theta_i(x,\tau) - \theta_j(x,\tau')). \tag{26}$$

Completely analogously as for $D$, one finds the recursion relations:

$$\frac{dD_m}{d\ln(s)} = (d + 2z - \frac{m^2}{Kc})D_m + O(D_n^2), \tag{27}$$

so that all $D_m$ with $m \geq 2$ are irrelevant at the SF-BG critical point, similarly to the irrelevance of vortices with higher vorticity at the Kosterlitz-Thouless transition of the 2D XY model.

## 5  Long-range Coulomb interaction

Now that the formalism is in place it is interesting to consider the case of long-range Coulomb interactions between bosons in the Hamiltonian (1), $v_c(r) = e^2/r$. The quadratic part of the long-distance action at $T = 0$ is now

$$S_{\phi,\Pi} = \int d^d\vec{x} d\tau (\frac{v_J}{2\pi}(\nabla \phi(\vec{x},\tau))^2 + \int d^d\vec{x}' \frac{e^2}{|\vec{x}-\vec{x}'|}\Pi(\vec{x},\tau)\Pi(\vec{x}',\tau) + i\Pi(\vec{x},\tau)\partial_\tau \phi(\vec{x},\tau)) \tag{28}$$

for $d > 1$. The canonical dimension of coupling $v_J$ thus remains the same as in the case of short-range interaction, while the canonical dimension of the charge $e$ can be read from the Eq. (28):

$$[e^2] = z - 1. \tag{29}$$

I define the long-range interaction in $d > 1$ as

$$v_c(\vec{r}) = \int \frac{d^d\vec{k}}{(2\pi)^d} \frac{e^2}{k^{d-1}} e^{i\vec{k}\cdot\vec{r}}. \tag{30}$$

Precisely in $d = 1$ this definition coincides with the short-range and not with the Coulomb interaction, which has the Fourier transform $\sim \ln(1/k)$ in $d = 1$. Since I am not interested in $d = 1$ per se, but only in $d > 1$, for the present purposes this distinction is not important [33]. The only difference in the scaling equations from the short-range case is then in the replacement of the recursion relation (10) with

$$\frac{dc}{d\ln(s)} = (z + \frac{d-3}{2} - \frac{4D}{K^2 c^4})c + O(D^2), \tag{31}$$

where $K$ is defined exactly as before, but $c^2 = 4v_J e^2$. The Eq. (31) may be understood as the recursion relation for charge, in place of the equation for the phonon velocity in the short-range case. Assuming that at the SF-BG fixed point the value of the charge is finite [28] fixes the value of $z$, by requiring $c = 1$ for example. Combining Eqs. (31), (9) and (11), similarly to the short-range case I find the dynamical exponent now to be

$$z = 1. \qquad (32)$$

This is again an exact result within the present formalism, for the same reason as in the short-range case. The correlation length exponent is

$$\nu = \frac{\sqrt{2}}{\sqrt{3\epsilon}} + O(\sqrt{\epsilon}). \qquad (33)$$

For $\epsilon = 1$ this yields $\nu \approx 0.82$, within the bounds set by the Monte Carlo calculations of Wallin et al. [20] $\nu = 0.90 \pm 0.15$. The inequality $\nu > 2/d$ is not satisfied by the lowest order estimate, but presumably higher-order terms would correct this. The anomalous dimension of the single-particle propagator is now

$$\eta = \frac{1}{3} - \frac{10\epsilon}{9} + O(\epsilon^2). \qquad (34)$$

The result for $\epsilon = 1$ is again much smaller than the Monte Carlo estimate [20] where $\eta = 0.8 \pm 0.4$. It may be interesting to note that the difference between anomalous dimensions for Coulomb and short-range interactions to the lowest order in $\epsilon$ is roughly the same as found numerically.

The analysis of the relevance of the residual interactions in the superfluid action from the preceeding section can easily be repeated in the present case with the same final conclusion. In fact, since one expects that $z = 1$ for the Coulomb interaction, both $v_x$ and $v_y$ appear irrelevant in $d = 2$ and $d = 3$ already by simple power counting.

## 6 Superfluid-Mott insulator transition in periodic potential

It is well known that in a purely periodic potential, Bose system at integer number of bosons per period undergoes a transition from a superfluid into an incompressible Mott insulator (MI) with increasing repulsion between bosons, which is in the universality class of the XY model [34]. Since the critical exponents of the XY model are known virtually exactly [35] one may test the presented method of calculation on this standard case. In $d = 1$ the effective action in Haldane representation becomes

$$S_\theta = 2\pi T_{xy} \int dx d\tau \{(\partial_x \theta(x,\tau))^2 + (\partial_\tau \theta(x,\tau))^2\} - v \int dx d\tau \cos(2\pi(\rho_0 - 1)x + 2\theta(x,\tau)), \qquad (35)$$

so that for one boson per period ($\rho_0 = 1$) it reduces to the sine-Gordon theory. Relativistic invariance of the theory implies immediately that $z = 1$. In $d = 1 + \epsilon$ the renormalization group equations for the SF-MI problem are simple deformations [36] of the celebrated Kosterlitz-Thouless recursion relations [37]:

$$\frac{du}{d\ln(s)} = -\epsilon(1-u) + 2v^2 + Buv^2 + O(v^4, u^2v^2) \tag{36}$$

$$\frac{dv}{d\ln(s)} = 2uv + Av^3 + O(uv^3, \epsilon v), \tag{37}$$

where $u = (1 - 1/8\pi T_{xy})$, $T_{xy}$ is the temperature of the equivalent classical XY model in $d = 2$ and $v$ is the fugacity of vortices. The coefficients in Eqs. (36) and (37) are known to be $B = 4$ and $A = -5$ [38], but I left general $A$ and $B$ in the recursion relations to emphasize that these two quantities, in contrast to the coefficients of the lowest order terms, separately are not universal. Linearization of the flow around the simple critical point of the recursion relations (36)-(37) then yields the correlation length exponent

$$\nu_{xy} = \frac{1}{2\sqrt{\epsilon}}(1 - \frac{2 + 2A + B}{8}\sqrt{\epsilon} + O(\epsilon)). \tag{38}$$

It is interesting that the exponent $\nu$ to this order depends only on the *universal* linear combination of the coefficients $A$ and $B$ [38]. This is a non-trivial check of the consistency of the expansion around the lower critical dimension for this problem, and suggests that the same universality may be expected in the SF-BG case as well. Taking $\epsilon = 1$ in the last equation one finds $\nu_{xy} = 0.75$, reasonably close to the $3D$ XY value of 0.67 [35]. For $\epsilon = 2$, $\nu_{xy} = 0.60$, somewhat farther from the expected Gaussian value $1/2$.

# 7 Summary and discussion

In this paper I have presented a new analytical approach to the critical behavior at the superfluid-Bose glass transition at $T = 0$ based on the expansion around the lower critical dimension $d = 1$. In the vicinity of $d = 1$ the SF-BG critical point becomes infinitesimal in disorder, which enables one to obtain the recursion relations for the coupling constants in the theory as expansions in small parameter $\epsilon = d - 1$. I found that the critical exponent is $z = d$ ($z = 1$) for the short-range (Coulomb) interaction between bosons, in accord with the conjectures based on finiteness of the compressibility [2] (charge [28]) at the transition. I obtained the correlation length exponent and anomalous dimension to two leading orders for the short-range and the Coulomb universality classes, and demonstrated that the analogous calculation for the superfluid-Mott insulator transition leads to the results in reasonable agreement with the known $\nu_{xy}$ exponent in $d = 2 + 1$. Higher-order anharmonic terms in the effective action for the superfluid which describe interactions between the excitations and the deviation of the dispersion relation from linearity are argued to be irrelevant at the SF-BG critical point, at least for $1 \leq d \leq 2$.

The critical point found in this work should be understood as describing the proliferation of topological defects in the superfluid ground state, which are *induced* by the disorder potential at $T=0$. In this sense the SF-BG (and even more SF-MI) transition is similar to Kosterlitz-Thouless finite temperature transition in the $XY$ model; both are defect-mediated transitions, except that the role of entropy term in the quantum ($T=0$) problem is played by the external (random or periodic) potential. This is precisely why the dual formulation of the problem is useful: it rewrites the theory in terms of "correct" degrees of freedom.

A similar expansion around one dimension for the superconductor-insulator transition has been attempted earlier by Kolomeisky [40], who argued that the physical critical point exists only in $d<1$, in contradiction with my conclusions. The mistake in his work lies in using the effective action in $d=1$ in Haldane form (7) to infer the canonical dimensions of the coupling constants in $d \neq 1$. The transformation from the particle density to the variable $\theta$ is a particular realization of the duality transformations [22], [30], [31], and in this form it is possible only in $d=1$ and meaningless everywhere else. For this reason one needs to determine the canonical dimensions of $K$ and $c$ from the effective action in the form (2), which is general and exists in any dimension. Note that this does not contradict my later use of the effective action in Eq. (7) to calculate the recursion relations, since the non-trivial parts of these are determined by the theory precisely in $d=1$.

The reader should note that the considered SF-BG transition in dirty bosonic system should be the right universality class for the s-wave superconductor - localized insulator transition in disordered electronic systems, at least in vicinity of one dimension. This is know to be true precisely in $d=1$ [24], and should remain true in its neighborhood as well, since the attractive interaction in the singlet channel will still open the gap in the spin sector [39].

Experimentally, the simplest universal quantity to measure is the product of dynamical and correlation length exponents, which follows for example from collapsing the finite temperature resistivity data on either insulating or superconducting side onto a universal curve [4]. Although the experimental situation is presently somewhat unclear, most measurements are consistent with $z \approx 2$ and $\nu \geq 1$ for the short-range interactions between bosons [4], [6]. It appears that before attempting a more realistic comparison with experiments the present calculation would need to be pushed to higher order. This may be done by remaining faithful to the logic of minimal substraction scheme which would require the higher order terms in the recursion relations to be calculated only in $d=1$. The results for the SF-MI transition in the periodic potential, where I relied on the already calculated higher order recursion relations, are encouraging in this sense.

The presented theory also facilitates a systematic treatment of number of interesting issues in the field, like the interplay between periodic potential and disorder [21], [41], the finite temperature crossovers close to the SF-BG criticality [42], and the corrections to scaling, which are left for future considerations.

It is a pleasure to acknowledge useful discussions with I. Affleck, M. Sharifzadeh Amin, V. Barzykin, P. Stamp, Z. Tešanović, M. Oshikawa and A. Zagoskin. The author is grateful to NSERC of Canada and the Izaak Walton Killam foundation for financial support and to Aspen center for physics where this work was initiated.

# 8 Appendix A: derivation of the effective action in $d = 1$

Haldane [27] proposed a way to improve upon the hydrodynamic theory in Eq. (2) in $d = 1$ by including the structure of the quantum fluid at shorter length scales. Here I outline the reasoning which leads to the effective action (7), which forms the basis for most of the calculations in this paper. Introduce a field $\theta''(x)$ so that $\partial_x \theta''(x) = \pi \rho(x)$. As the density is discreet and therefore a sum of delta-functions, $\theta''(x)$ jumps by $\pi$ at the positions of particles. So we may write

$$\rho(x) = \pi^{-1} \partial_x \theta''(x) \sum_{n=-\infty}^{\infty} \delta[\theta''(x) - n\pi], \tag{39}$$

or using the Poisson summation formula and introducing the local deviation from the average density $\Pi(x)$:

$$\rho(x) = (\rho_0 + \Pi(x)) \sum_{m=-\infty}^{\infty} e^{i2m\theta''(x)}. \tag{40}$$

The $m = 0$ term in the last formula may then be understood as the coarse-grained density, and the rest of the sum yields shorter and shorter length-scale corrections. The imaginary time action in Eq. (2) together with the random potential term now becomes

$$S_{\phi,\Pi} = \int dx \int_0^\beta d\tau (\frac{v_J}{2\pi} (\partial_x \phi(x,\tau))^2 + 2\pi v_N \Pi^2(x,\tau) + i\Pi(x,\tau) \partial_\tau \phi(x,\tau) + \tag{41}$$
$$V(x)\Pi(x,\tau) + \rho_0 V(x)(e^{i2\theta''(x,\tau)} + c.c.)) + hh$$

where "hh" denotes the higher harmonics that arise from the infinite sum in Eq. (40). Introducing a new field $\theta'(x,\tau)$ as $\pi^{-1} \partial_x \theta'(x,\tau) = \Pi(x,\tau)$, one may write the term that couples the particle density and the superfluid phase as

$$\frac{i}{\pi} \int dx \int_0^\beta d\tau (\partial_x \phi(x,\tau))(\partial_\tau \theta'(x,\tau)). \tag{42}$$

The phase $\phi$ can now be straightforwardly integrated out which leads to

$$S_\theta = \int dx \int_0^\beta d\tau (\frac{2v_N}{\pi} (\partial_x \theta'(x,\tau))^2 + \frac{1}{2\pi v_J} (\partial_\tau \theta'(x,\tau))^2 + \tag{43}$$
$$V(x) \frac{1}{\pi} \partial_x \theta'(x,\tau) + \rho_0 V(x)(e^{2i(\theta'(x,\tau)+\pi\rho_0 x)} + c.c.)).$$

Fourier components of $V(x)$ with the wave-vectors $k << \rho_0$ couple only to $\Pi(\vec{x})$ in Eq. (43), and can be completely absorbed by a shift of $\Pi$, i. e. by defining a yet another field $\theta$ as in Eq. (8). These components lead only to forward scattering, and in $d = 1$ are unrelated to localization. The components of $V(x)$ with $k \sim 2\pi\rho_0$ however couple directly to $\exp i2\theta'(x)$, which is the operator that creates $2\pi$ phase-slips in the ground state, and can not be eliminated this way. Assuming the Gaussian probability distribution

$$P[V(k)] \sim e^{-\frac{V^*(k)V(k)}{D}} \tag{44}$$

for the Fourier components with $k \sim 2\pi\rho_0$, after introducing replicas and averaging over $V(k)$ one obtains the Eq. (7) as in the text.

For a different, and maybe somewhat more rigorous derivation of the theory (7) the reader should consult the second paper in the reference [21].

## 9 Appendix B: momentum shell RG

Define the slow and the fast components of the field $\theta$:

$$\theta(x,\tau) = \theta_1(x,\tau) + \theta_2(x,\tau), \tag{45}$$

where

$$\theta_1(x,\tau) = \int_{-\Lambda/s}^{\Lambda/s} \frac{dk}{2\pi} \int_{-\infty}^{\infty} \frac{d\omega}{2\pi} \theta(k,\omega) e^{ikx+i\omega\tau}, \tag{46}$$

and analogously the $\theta_2$, but with the integral over momenta going over $\Lambda/s < |k| < \Lambda$. To the lowest order in $D$, after integrating out the fast modes the remaining action for the slow modes becomes

$$S_{\theta_1} = K \sum_i \int dx d\tau (c^2 (\partial_x \theta_{1,i}(x,\tau))^2 + (\partial_\tau \theta_{1,i}(x,\tau))^2) - \tag{47}$$

$$D\rho_0^2 \sum_{i,j} \int dx d\tau d\tau' \cos 2(\theta_{1,i}(x,\tau) - \theta_{1,j}(x,\tau'))\langle \cos 2(\theta_{2,i}(x,\tau) - \theta_{2,j}(x,\tau'))\rangle_2 + O(D^2),$$

where the average is performed over the quadratic part of the action for the fast modes. First, consider the terms off-diagonal in replica indices. One finds

$$\langle \cos 2(\theta_{2,i}(x,\tau) - \theta_{2,j}(x,\tau'))\rangle_2 = s^{-\frac{1}{\pi Kc}} \tag{48}$$

for $i \neq j$. If $i = j$

$$\langle \cos 2(\theta_{2,i}(x,\tau) - \theta_{2,i}(x,\tau'))\rangle_2 = \exp -(\frac{\Lambda \ln(s)}{\pi^2 K} \int_{-\infty}^{\infty} \frac{d\omega}{c^2\Lambda^2 + \omega^2}(1 - e^{i\omega(\tau-\tau')}) + O(\ln(s)^2)). \tag{49}$$

The first term in the bracket in the $O(\ln(s))$-term matches the contribution from the off-diagonal terms in Eq. (48) and together they renormalize $D$:

$$D \to D(s) = Ds^{(3-(\pi Kc)^{-1})} + O(D^2). \tag{50}$$

After expanding the cosine-term in $S_{\theta_1}$ to the quadratic order we see that the time-dependent term in Eq. (49) apart from generating some irrelevant terms also renormalizes the coefficient in front of $(\partial_\tau \theta_1)^2$. Thus, for an infinitesimal $\ln(s)$

$$K \to K(s) = K + \frac{8D\rho_0^2}{\pi K c^4 \Lambda^3} \ln(s) + O(D^2, \ln(s)^2). \tag{51}$$

To the first order in $D$ the non-local term in $x$ does not get corrected by the elimination of the slow modes:

$$Kc^2 \to K(s)c(s)^2 = Kc^2 + O(D^2). \tag{52}$$

After redefinitions of the coupling as described right above the Eq. (9) we see that differentiating the above expressions yields the lowest-order terms in the Eqs. (9)-(11) when $d = z = 1$.

Next, consider the contribution quadratic in $D$ to the $S_{\theta_1}$:

$$-\frac{D^2 \rho_0^4}{2} \sum_{i,j,m,n} \int dx d\tau d\tau' dy dv dv' \{\cos 2(\theta_{1,i}(x,\tau) - \theta_{1,j}(x,\tau')) \cos 2(\theta_{1,m}(y,v) - \theta_{1,n}(y,v')) \tag{53}$$

$$[\langle \cos 2(\theta_{2,i}(x,\tau) - \theta_{2,j}(x,\tau')) \cos 2(\theta_{2,m}(y,v) - \theta_{2,n}(y,v')) \rangle_2 -$$
$$\langle \cos 2(\theta_{2,i}(x,\tau) - \theta_{2,j}(x,\tau')) \rangle_2 \langle \cos 2(\theta_{2,m}(y,v) - \theta_{2,n}(y,v')) \rangle_2 ] +$$
$$\sin 2(\theta_{1,i}(x,\tau) - \theta_{1,j}(x,\tau')) \sin 2(\theta_{1,m}(y,v) - \theta_{1,n}(y,v'))$$
$$\langle \sin 2(\theta_{2,i}(x,\tau) - \theta_{2,j}(x,\tau')) \sin 2(\theta_{2,m}(y,v) - \theta_{2,n}(y,v')) \rangle_2 \}.$$

Using the identities:

$$\langle \cos \alpha \cos \beta \rangle_2 = e^{-\frac{\langle \alpha^2 \rangle_2 + \langle \beta^2 \rangle_2}{2}} \cosh \langle \alpha \beta \rangle_2, \tag{54}$$

$$\langle \sin \alpha \sin \beta \rangle_2 = e^{-\frac{\langle \alpha^2 \rangle_2 + \langle \beta^2 \rangle_2}{2}} \sinh \langle \alpha \beta \rangle_2, \tag{55}$$

and realizing that all averages on the right hand side in the last two equations are $\sim \ln(s)$, we see that the first term in Eq. (53) is of order $(\ln(s))^2$. To the first order in $\ln(s)$ only the second term in Eq. (53) can potentially contribute to the action for the remaining slow modes, and it equals to

$$-8D^2 \rho_0^4 \sum_{i,j,n} \int dx d\tau d\tau' dy dv dv' \sin 2(\theta_{1,i}(x,\tau) - \theta_{1,j}(x,\tau')) \tag{56}$$

$$\sin 2(\theta_{1,i}(y,v) - \theta_{1,n}(y,v')) \langle \theta_{2,i}(x,\tau) \theta_{2,i}(y,v) \rangle_2 + O((\ln(s))^2).$$

Since the corelator appearing in the last equation drops exponentially in $|\tau - v|$ and $|x - y|$, the dominant contribution comes from the leading term in the expansion of the fields under the second sinus around the point $(x, \tau)$ and $(x, \tau')$. For $n = j$ one finds the term

$$-\frac{8D^2 \rho_0^4 \ln(s)}{4\pi K c \Lambda^3} \sum_{i,j} \int dx d\tau d\tau' \sin^2 2(\theta_{1,i}(x,\tau) - \theta_{1,j}(x,\tau')), \tag{57}$$

which obviously, besides adding a constant to the free energy, just generates a higher harmonic. The recursion relation in Eq. (11) is therefore correct to $O(D^3)$. There may be $O(D^2)$ renormalization of $K$ deriving from the term (56), but the reader may convince himself that this does not matter for the exponents $\nu$ and $\eta$ to the two lowest orders in $\epsilon$.

# 10 Appendix C: proof that $z = d$ to all orders

I derive the general form of the recursion relation for the coupling constant $c$, from which it will follow that for the short-range interactions $z = d$ within the considered renormalization procedure. First step is to show that the coefficient in front of the $(\partial_x \theta)^2$- term in the theory (7) in $d = 1$, $Kc^2 = 2v_N/\pi$, can not become renormalized during the integration over the fast modes. Let me assume that the Bose system is of a finite size $L$ and impose the periodic boundary conditions in space direction:

$$\theta_i(0, \tau) = \theta_i(L, \tau) \tag{58}$$

for all $0 < \tau < 1/T$. To be precise I will assume a finite temperature $T$, although this does not affect the argument in any way. The renormalized $v_N(L)$ at the scale $L$ can be defined as

$$v_N(L) = \frac{\pi k_B T L}{2\hbar} \frac{d\overline{F}(\delta)}{d(\delta^2)}\Big|_{\delta=0}, \tag{59}$$

where $\delta$ is the imposed twist in the boundary conditions in the space direction:

$$\theta_i(L, \tau) = \theta_i(0, \tau) + \delta. \tag{60}$$

Defining a new field $\theta'_i(x, \tau) = \theta_i(x, \tau) - \delta x/L$ which satisfies the periodic boundary conditions, and noticing that the disorder term in Eq. (7) is invariant under this transformation, it readily follows that

$$v_N(L) = v_N, \tag{61}$$

independently of the length scale $L$. In other words, under the scale transformation $x \to sx$

$$\frac{d}{d\ln(s)} Kc^2 = 0 \tag{62}$$

exactly, in $d = 1$. If one defines a function $F(D, K, c)$ which is not explicitly dependent on dimension as

$$\frac{dK}{d\ln(s)} = (2 - z - d)K + F(D, K, c) \tag{63}$$

in $d > 1$, then Eq. (62) implies that

$$\frac{dc}{d\ln(s)} = (z - 1 - \frac{F(D, K, c)}{2K})c. \tag{64}$$

Demanding that the last two renormalization group equations vanish at some non-trivial couplings $D^*$, $K^*$ and $c^*$, implies that at the fixed point

$$z = d. \tag{65}$$

# 11   Appendix D: Bose correlation function at the criticality

It is worth showing how the algebraic decay of the disorder averaged single-particle correlation function at the critical point arises in $d > 1$. The Bose operator correlation function averaged over disorder is

$$\overline{G(\vec{x})} \sim \rho_0 \langle e^{i\phi_1(\vec{x},0) - i\phi_1(0,0)} \rangle, \tag{66}$$

where the average is to be taken over the replicated action. Right at the SF-BG transition, the above average becomes

$$\overline{G(\vec{x})} \sim \rho_0 \exp -2\pi K^* c^* \int \frac{d^{d+z}\vec{k}}{(2\pi)^{d+1}} \frac{1 - e^{i\vec{k}\cdot\vec{x}}}{k^{2-\eta_\phi}}, \tag{67}$$

where $\eta_\phi$ is the anomalous dimension of the phase corelator, defined as

$$\eta_\phi = \frac{d \ln Z_K}{d \ln(k)}, \tag{68}$$

where the renormalization factor $Z_K$ is defined as $K(s) \equiv K Z_K$. Since under the scale transformation $\Lambda \to \Lambda/s$, $k \to k/s$, it follows that

$$\eta_\phi = -\frac{d \ln Z_K}{d \ln(s)}. \tag{69}$$

The recursion relation for K may then by definition be written as

$$\frac{dK}{d \ln(s)} = (2 - z - d - \eta_\phi)K, \tag{70}$$

so that at the fixed point we must have $\eta_\phi = 2 - z - d$. Inserting this into Eq. (67) the expression in the exponential becomes

$$\frac{K^* c^*}{(2\pi)^{d-1}} (\prod_{n=2}^{2d-2} \int_0^\pi \sin^n \theta d\theta) \int_0^\Lambda \frac{dk}{k} \int_0^\pi \sin \theta (1 - e^{ikx \cos \theta}) d\theta \to \frac{K^* c^* \ln(\Lambda x)}{2^\epsilon \Gamma(1+\epsilon)}, \tag{71}$$

for $x \to \infty$, and I assumed $d = z$. The explicit $\epsilon$-dependence in the result arises from the integration in $d > 1$. Adhering to the philosophy of minimal substraction we should set $\epsilon = 0$ in the last equation, so that the only dependence on dimensionality is the one implicit in the value of $K^*$ and $c^*$ at the fixed point. Thus Eq. (17) obtains.

Figure captions:

Figure 1: The renormalization group flow in $d > 1$. Close to $d = 1$ the fixed point which controls the transition between the superfluid ($D \to 0$, $K \to 0$) and the localized ($D \to \infty$, $K \to \infty$) phases is at a small value of disorder $D^* \sim \epsilon$. Inset a) depicts the flow in $d < 1$, where only the localized phase is stable, and the repulsive fixed point lies in the unphysical region $D < 0$. Precisely at $d = 1$ the fixed point lies on the marginal line at $D = 0$ (inset b).

# References

[1] M. Ma and P. A. Lee, Phys. Rev. B **32**, 5658 (1985); M. Ma, B. I. Halperin, and P. A. Lee, Phys. Rev. B **34**, 3136 (1986).

[2] M. P. A. Fisher, P. B. Weichman, G. Grinstein and D. S. Fisher, Phys. Rev. B **40**, 546 (1989).

[3] H. S. J. van der Zant, W. J. Elion, L. J. Geerlings, and J. E. Mooij, Phys. Rev. B **54**, 10081 (1996) and the references therein.

[4] Y. Liu and A. M. Goldman, Mod. Phys. Lett. B **8**, 277 (1994); J. M. Valles Jr., R. C. Dynes and J. P. Garno, Phys. Rev. Lett. **69**, 3567 (1992); A. Yazdani and A. Kapitulnik, Phys. Rev. Lett. **74**, 3037 (1995).

[5] S. Doniach and M. Inui, Phys. Rev. B **41**, 6668 (1990); Y. Fukuzumi, K. Mizuhashi, K. Takenaka, and S. Uchida, Phys. Rev. Lett. **76**, 684 (1996).

[6] P. A. Crowell, F. W. van Keuls, and J. D. Reppy, Phys. Rev. Lett. **75**, 1106 (1995); Phys. Rev. B **55**, 12620 (1997).

[7] J. V. Porto and J. M. Parpia, Phys. Rev. Lett. **74**, 4667 (1995); K. Matsumoto, J. V. Porto, L. Pollack, E. N. Smith, T. L. Ho, and J. M. Parpia, Phys. Rev. Lett. **79**, 253 (1997).

[8] S. V. Kravchenko, W. E. Mason, G. E. Bowker, J. E. Furneaux, V. M. Pudalov and M. D'Iorio, Phys. Rev. B **51**, 7038 (1995); V. Dobrosavljević, E. Abrahams, E. Miranda, and S. Chakravarty, Phys. Rev. Lett. **79**, 455 (1997).

[9] D. R. Nelson and V. Vinokur, Phys. Rev. Lett. **68**, 2398 (1992); Phys. Rev. B **48**, 13060 (1993); W. Jiang, N. C. Yeh, D. S. Reed, U. Kriplani, D. A. Beam, M. Konczykowski, T. A. Tombrello, and F. Holtzberg, Phys. Rev. Lett **72**, 550 (1994).

[10] Z. Tešanović and I. F. Herbut, Phys. Rev. B **50**, 10389 (1994); I. F. Herbut, in *Fluctuation phenomena in high temperature superconductors*, ed. by M. Ausloos and A. Varlamov (Kluwer, Dordrecht, 1997), p. 311.

[11] See, for example S. K. Ma, *Modern theory of critical phenomena*, (Benjamin, Reading, MA, 1976).

[12] T. C. Lubensky, Phys. Rev. B **11**, 3573 (1975); D. E. Khmelnitskii, Phys. Lett. **67** A, 59 (1978); A. Weinrib and B. I. Halperin, Phys. Rev. B **27**, 413 (1983).

[13] S. N. Dorogovtsev, Phys. Lett. A **76**, 169 (1980); D. Boyanovski and J. Cardy, Phys. Rev. B **26**, 154 (1982); I. D. Lawrie and V. V. Prudnikov, J. Phys. C **17**, 1655 (1984).

[14] A. B. Harris, J. Phys. C **7**, 1671 (1974); J. T. Chayes, L. Chayes, D. S, Fisher, and T. Spencer, Phys. Rev. Lett. **57**, 2999; (1986); See however, F. Pazmandi, R. T. Scalettar, and G. T. Zimanyi, Phys. Rev. Lett. **79**, 5130 (1997).

[15] P. B. Weichman and K. Kim, Phys. Rev. B **40**, 813 (1989).

[16] See also, R. Mukhopadhyay and P. B. Weichman, Phys. Rev. Lett. **76**, 2997 (1996).

[17] Y. Tau and P. B. Weichman, Phys. Rev. Lett. **73**, 6 (1994); Y. B. Kim and X. G. Wen, Phys. Rev. B **49**, 4043 (1994).

[18] L. Zhang and M. Ma, Phys. Rev. B **45**, 4855 (1992); K. Singh and D. Rokshar, Phys. Rev. B **46**, 3002 (1992).

[19] J. Freericks and H. Monien, Phys. Rev. B **53**, 2691 (1996).

[20] K. Runge, Phys. Rev. B **45**, 13136 (1992); G. G. Batrouni, B. Larson, R. T. Scalettar, J. Tobochnik, and J. Wang, Phys. Rev. B **48**, 9628 (1993); M. Makivić, N. Trivedi, and S. Ullah, Phys. Rev. Lett. **71**, 2307 (1993); M. Wallin, E. Sorensen, S. M. Girvin and A. P. Young, Phys. Rev. B **49**, 12115 (1994) and references therein.

[21] I. F. Herbut, Phys. Rev. Lett. **79**, 3502 (1997); unpublished, preprint cond-mat/9801083.

[22] I. F. Herbut, Phys. Rev. B **57**, 1303 (1998).

[23] W. Apel, J. Phys. C: Solid State Phys. **15**, 1973 (1982).

[24] T. Giamarchi and H. J. Shulz, Phys. Rev. B **37**, 327 (1988).

[25] N. D. Mermin and H. Wagner, Phys. Rev. Lett. **22**, 1133 (1966); P. C. Hohenberg, Phys. Rev. B **158**, 383 (1967).

[26] V. N. Popov, *Functional integrals in quantum field theory and statistical mechanics* (Reidel, Boston, 1983), Chapter 6.

[27] F. D. M. Haldane, Phys. Rev. Lett. **51**, 605 (1983).

[28] M. P. A. Fisher and G. Grinstein, Phys. Rev. Lett. **60**, 208 (1988); M. P. A. Fisher, G. Grinstein and S. M. Girvin, Phys. Rev. Lett. **64**, 587 (1990).

[29] I. M. Lifshitz and L. P. Pitaevskii, *Statistical Physics* (Pergamon, New York, 1970), chapter 3.

[30] M. P. A. Fisher and D. H. Lee, Phys. Rev. B **39**, 2756 (1989).

[31] I. F. Herbut, J. Phys. A: Math. Gen. **30**, 423 (1997).

[32] G. Grinstein, in *Fundamental problems in statistical mechanics VI*, ed. by E. G. D. Cohen, (North-Holland, Amsterdam, 1985), p. 147.

[33] In $d = 1$ real Coulomb interaction completely suppresses superfluidity, and the ground state is always insulating. See ref. [28], or more recently: S. R. Renn and J. M. Duan, Phys. Rev. Lett. **76**, 3400 (1996).

[34] S. Doniach, Phys. Rev. B **24**, 5063 (1981).

[35] J. Zinn-Justin, *Quantum Field Theory and Critical Phenomena* (Oxford University Press, Oxford, 1993).

[36] J. L. Cardy and H. W. Hamber, Phys. Rev. Lett. **45**, 499 (1980); D. R. Nelson and D. S. Fisher, Phys. Rev. B **16**, 4945 (1977).

[37] J. M. Kosterlitz and D. J. Thouless, J. Phys. C **6**, 1181 (1973); J. M. Kosterlitz, J. Phys. C **7**, 1046 (1974).

[38] D. J. Amit, Y. Y. Goldschmidt and G. Grinstein, J. Phys. A: Math. Gen. **13**, 585 (1980).

[39] K. Ueda and T. M. Rice, Phys. Rev. B, **29**, 1514 (1984); see also ref. [22].

[40] E. B. Kolomeisky, Phys. Rev. B **48**, 4998 (1993).

[41] W. Krauth, N. Trivedi, D. Ceperly, Phys. Rev. Lett. **67**, 2307 (1991); R. V. Pai, R. Pandit, H. R. Krishnamurthy, and S. Ramasesha, Phys. Rev. Lett. **76**, 2937 (1996); B. Svistunov, Phys. Rev. B **54**, 16131 (1996); J. Kisker and H. Rieger, Phys. Rev. B **55**, R11981 (1997); F. Pazmandi, G. T. Zimanyi, and R. Scalettar, Phys. Rev. Lett. **75**, 1356 (1995); F. Pazmandi and G. T. Zimanyi, unpublished, preprint cond-mat/9702169.

[42] K. Damle and S. Sachdev, Phys. Rev. B **56**, 8714 (1997), S. Sondhi, S. M. Girvin, J. P. Carini and D. Shahar, Rev. Mod. Phys. **69**, 315 (1997).

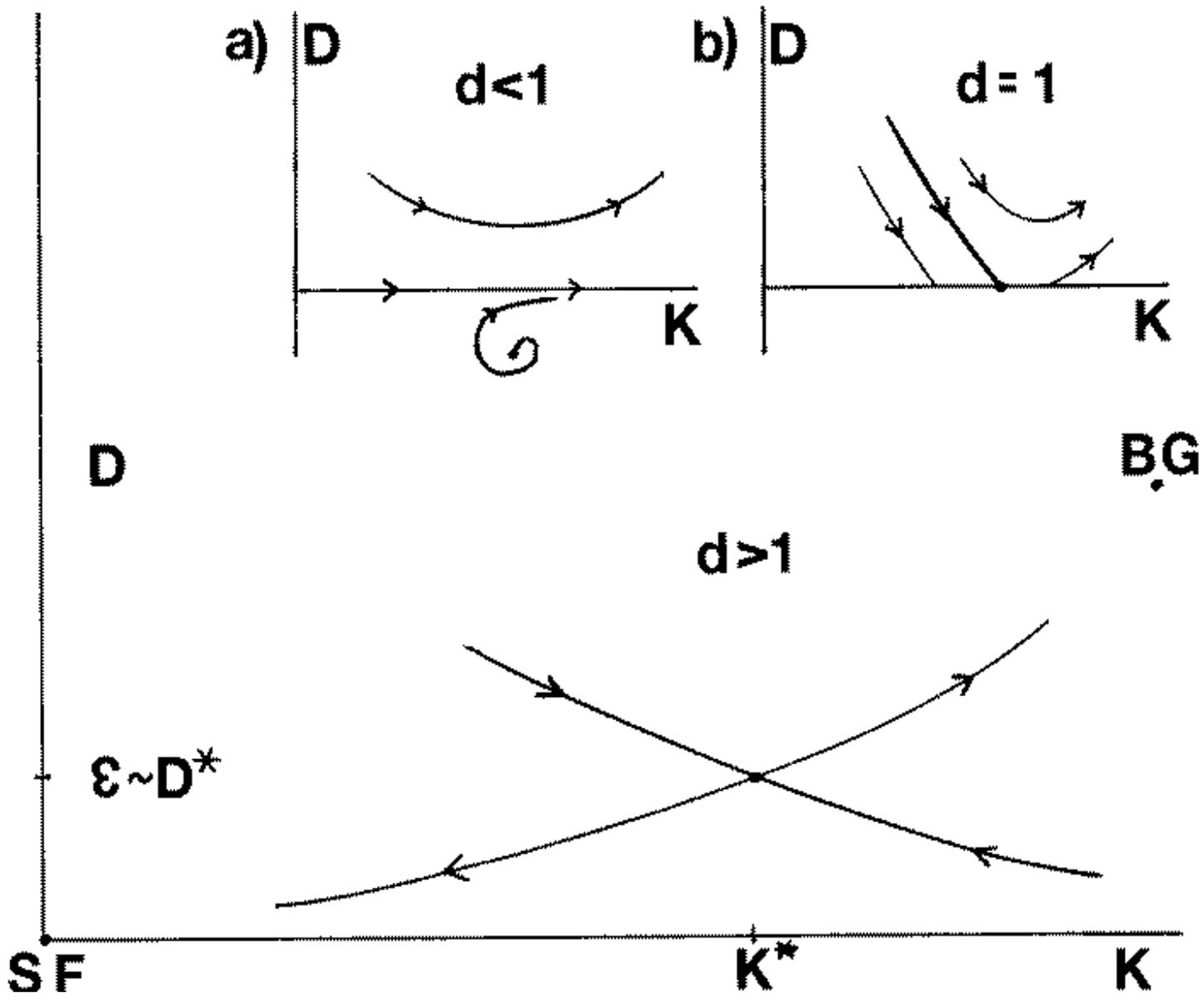